\documentclass[english,preprint,12pt]{elsarticle}

\usepackage[T1]{fontenc}
\usepackage[latin1]{inputenc}
\usepackage{graphicx,here,upgreek}
\usepackage{epsfig}
\usepackage{lineno}
\usepackage{amssymb}
\usepackage{longtable}
\usepackage{xspace}
\def\Offline{\mbox{$\overline{\rm 
Off}$\hspace{.05em}\raisebox{.3ex}{$\underline{\rm line}$}}\xspace}

\makeatother

\begin{document}

\begin{frontmatter}

\title{Trigger and Aperture of the Surface Detector Array of the Pierre Auger Observatory}

\author{
{\bf The Pierre Auger Collaboration} \\
J.~Abraham$^{8}$, 
P.~Abreu$^{71}$, 
M.~Aglietta$^{54}$, 
C.~Aguirre$^{12}$, 
E.J.~Ahn$^{87}$, 
D.~Allard$^{31}$, 
I.~Allekotte$^{1}$, 
J.~Allen$^{90}$, 
J.~Alvarez-Mu\~{n}iz$^{78}$, 
M.~Ambrosio$^{48}$
% ...
}
\address{
\par\noindent
$^{1}$ Centro At\'{o}mico Bariloche and Instituto Balseiro (CNEA-
UNCuyo-CONICET), San Carlos de Bariloche, Argentina \\
% ...
$^{8}$ National Technological University, Faculty Mendoza 
(CONICET/CNEA), Mendoza, Argentina \\
% ...
$^{12}$ Universidad Catolica de Bolivia, La Paz, Bolivia \\
% ...
$^{31}$ Laboratoire AstroParticule et Cosmologie (APC), 
Universit\'{e} Paris 7, CNRS-IN2P3, Paris, France \\
% ...
$^{48}$ Universit\`{a} di Napoli ``Federico II'' and Sezione INFN, 
Napoli, Italy \\
% ...
$^{54}$ Istituto di Fisica dello Spazio Interplanetario (INAF),
 Universit\`{a} di Torino and Sezione INFN, Torino, Italy \\
% ...
$^{71}$ LIP and Instituto Superior T\'{e}cnico, Lisboa, Portugal \\
% ...
$^{78}$ Universidad de Santiago de Compostela, Spain \\
% ...
$^{87}$ Fermilab, Batavia, IL, USA \\
% ...
$^{90}$ New York University, New York, NY, USA \\
% ...
(\ddag) Deceased \\
(a) at Konan University, Kobe, Japan \\
(b) On leave of absence at the Instituto Nacional de Astrofisica, Optica y Electronica \\
(c) at Caltech, Pasadena, USA \\
(d) at Hawaii Pacific University \\
}

\cortext[cor1]{Corresponding author. Email address: giorgio.matthiae@roma2.infn.it}

\begin{abstract}

The surface detector array of the Pierre Auger Observatory consists of 1600 water-Cherenkov detectors, for the study of extensive air showers (EAS) generated by ultra-high-energy cosmic rays.  
We describe the trigger hierarchy, from the identification of candidate showers at the level of a single detector, amongst a large background (mainly random single cosmic ray muons), up to the selection of real events and the rejection of random coincidences. Such trigger makes
 the surface detector array  fully efficient for the detection of EAS with energy above $3\times 10^{18}$ eV, for all zenith angles between 0$^\circ$ and 60$^\circ$,  independently of the position of the impact point and of the mass of the primary particle. In these range of energies and angles, the exposure of the surface array can be determined purely on the basis of the geometrical acceptance.

\end{abstract}


\end{frontmatter}
\linenumbers
\section{Introduction}
The main objective of the Pierre Auger Collaboration is to
measure the flux, arrival direction distribution and mass
composition of cosmic rays from $\approx 10^{18}$ eV up to the
highest energies. Due to the very low fluxes at these energies,
cosmic rays have to be measured through the \emph{extensive air
showers} (EAS) they produce in the atmosphere.

The Pierre Auger Observatory, located near Malarg\"ue, Argentina, at 1400 m asl, detects EAS in two independent and complementary ways. It includes a \emph{surface detector array} (SD), consisting of 1600 water-Cherenkov detectors \cite{SDpaper} on a triangular grid of 1.5 km spacing covering an area of approximately 3000 km$^2$, which detects the secondary
particles at ground level and thus samples their lateral density
distribution. The surface detector array is overlooked by a \emph{fluorescence detector} (FD) consisting of 24 telescopes at four sites, which measure the
fluorescence light emitted along the path of the air-showers and
thus traces their longitudinal development \cite{FDpaper}. Showers detected by both detectors are called \emph{hybrid events} and they are characterised more accurately with respect to direction and energy than using either technique alone. However, the livetime of the FD is limited to $\approx$
13\%, as it only operates on clear, moonless nights \cite{FDpaper}. The bulk of data
is provided by the SD with its nearly 100\% livetime.
The study of the trigger and the determination of the aperture of the SD is thus essential for
the physics aims of the Pierre Auger Observatory.

The SD data acquisition (DAQ) trigger must fulfill  both physical and technical requirements. The main limitation to the rate of recordable events comes from the wireless communication system which connects the surface detectors to the central campus. The latter must serve continuously 1600 stations spread over 3000 km$^2$,  each using an emitter consuming less than 1 W power to transmit to collectors as far as 40 km away.  The maximum sustainable rate of events per detector is less than one per hour, to be compared to the 3 kHz counting rate per station, due to the atmospheric muon flux. The trigger thus must reduce the single station rate, without inducing loss of physics events. It must also allow data acquisition down to the lowest possible energy. To deal with all these requirements, the design of the DAQ SD trigger (described in
section 3) has been realised in a hierarchical form, where at each
level the single station rate becomes less and less, by means of discrimination against background stricter and stricter. At the same time, the DAQ trigger is designed to allow the storage of the largest possible number of EAS  candidates .

The ultimate discrimination of EAS from chance events due to combinatorial coincidences among the surface detectors is performed off-line through a selection of physics events, and of detectors participating in each of them. The event selection procedure is hierarchical too, it is described in section 4.

In section 5.1, we show that the trigger and event selection hierarchy makes the array fully efficient for the detection of showers above  $3\times 10^{18}$ eV. We restrict ourselves to this energy range for the calculation of the exposure
(described in section 5.2), which is simply proportional to the observation
time and to the geometrical size of the SD array. Under these conditions the
calculation of the exposure is very robust and almost devoid of
systematic uncertainties. Therefore it is straightforward to calculate the cosmic ray flux as the ratio of the number of collected events to the effective, as it was done in the measurement of the cosmic ray spectrum by the surface detector of Auger \cite{spectrum}. 

\section{The surface detector of the Pierre Auger Observatory}
Each water Cherenkov detector of the surface array has a 10 $\textrm{m}^{2}$ water surface
area and 1.2 m water depth, with three 9'' photomultiplier tubes (PMTs) looking through optical coupling material into the water volume, which is contained in a Tyvek$^{\textregistered}$
reflective liner \cite{SDpaper,auger_nim}. Each detector operates
autonomously, with its own electronics and communications systems
powered by solar energy. Each PMT provides two signals, which are digitised by 40
MHz 10-bit Flash Analog to Digital Converters (FADCs). One signal is
directly taken from the anode of the PMT, and the other signal is
provided by the last dynode, amplified and inverted within the PMT base
electronics to a total signal nominally 32 times the anode
signal. The two signals are used to provide sufficient dynamic range to
cover with good precision both the signals produced in the detectors near the shower core ($\sim
1000~\textrm{particles}/\textrm{$\mu s$}$) and those produced far
from the shower core ($\sim 1~\textrm{particle}/\textrm{$\mu s$}$).
Each FADC bin corresponds to 25 ns \cite{auger_nim}.\\
The signals from the three PMTs are sent to a central data acquisition system (CDAS) once a candidate shower event triggers the surface detector array (see section 3.2). The total bandwidth available for data transmission from the detectors to the CDAS is 1200 bits per second, which precludes the possibility of any remote calibration. For this reason, the calibration of each detector is performed locally and automatically. It relies on the measurement of the average charge collected by a PMT from the Cherenkov light produced by a vertical and central through-going muon, $Q_{VEM}$ \cite{CalibrationPaper}. The water-Cherenkov detector in its normal configuration has no way to select only vertical muons. However, the distribution of the light of atmospheric muons produces a peak in the charge distribution, $Q^{peak}_{VEM}$ (or VEM in short), as well as a peak in that of the pulse height, $I^{peak}_{VEM}$, both of them being proportional to those produced by a vertical through-going muon. The calibration parameters are determined with 2\% accuracy every 60 s and returned to the CDAS with each event.
Due to the limited bandwidth, the first level triggers are also performed locally. These triggers (section~3.1) are set in electronic units (channels): the reference unit is $I^{peak}_{VEM}$.

With respect to shower reconstruction, the signals recorded by the detectors - evaluated by integrating the FADC bins of the traces - are converted to units of $Q_{VEM}$. These are fitted with a measured Lateral Distribution Function (LDF) \cite{LDF}, that describes $S(r)$, the signals as a function of distance $r$ from the shower core, to find the signal at 1000 m, $S$(1000) \cite{s1000}. The variation of $S$(1000) with zenith angle $\uptheta$ arising from the evolution of the shower, is quantified by applying the constant integral intensity cut method \cite{CIC}, justified by the approximately isotropic flux of primary cosmic rays. An energy estimator for each event, independent of $\uptheta$, is $S_{38}$, the $S$(1000) that EAS would have produced had they arrived at the median zenith angle, $38^\circ$. The energy corresponding to each $S_{38}$ is then obtained through a calibration with the fluorescence detector based on a subset of high-quality hybrid events \cite{spectrum}.

\section{The DAQ trigger system of the surface detector array}
\label{sec:trigger}
\begin{figure}[H]
\centering
\includegraphics[width=14cm]{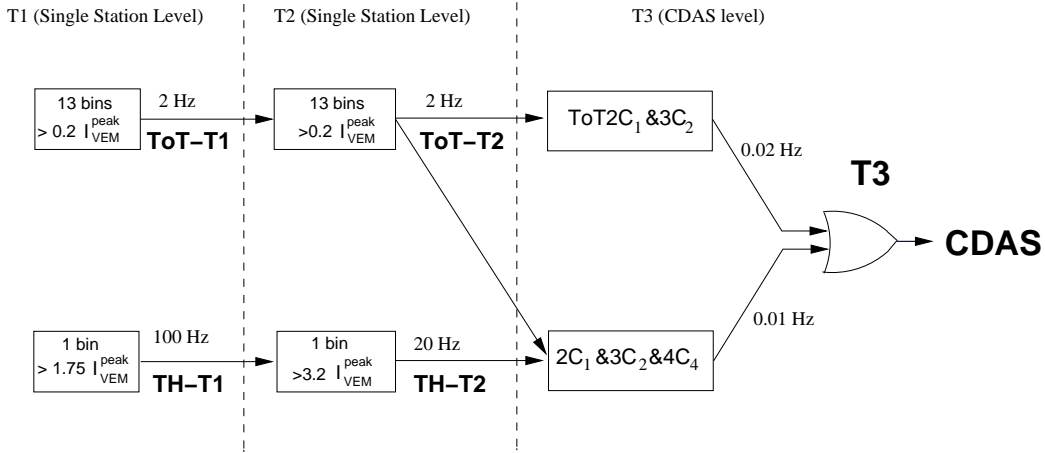}
\caption{Schematics of the hierarchy of the trigger system of the Auger surface detector.}
\label{fig:schema}
\end{figure}

The trigger for the surface detector array is hierarchical. Two levels of trigger (called T1 and T2) are formed at each detector.  T2 triggers are combined with those from other detectors and examined for spatial and temporal correlations, leading to an array trigger (T3).  The T3 trigger initiates data acquisition and storage. The logic of this trigger system is summarised in figure~\ref{fig:schema}.

\subsection{Single detector triggers}\label{ls}
The {\bf T1} triggers data acquisition in each water Cherenkov detector: data are stored on the local disk for 10 s waiting for a possible T3. Two independent trigger modes are implemented as T1, having been conceived to detect, in a complementary way, the electromagnetic and muonic components of an air-shower.  The first T1 mode is a simple threshold trigger (TH) which requires the coincidence of the three PMTs each above 1.75 $I^{peak}_{VEM}$\footnote{For detectors with only two (one) operating PMTs the threshold is 2 (2.8) $I^{peak}_{VEM}$.}. This trigger is used to select large signals that are not necessarily spread in time.  It is particularly effective for the detection of very inclined showers that have crossed a large amount of atmosphere and are consequently dominantly muonic.  The TH-T1 trigger is used to reduce the rate due to atmospheric muons from $\approx$3 kHz to $\approx$100 Hz.  The second T1 mode makes use of the fact that, for other than very inclined showers or signals from more vertical showers very close to the shower axis, the arrival of particles and photons at the detector is dispersed in time  \cite{Linsley,Watson}.  For example, at 1000 m from the axis of a vertical shower, the time for the signal from a water-Cherenkov detector to rise from 10 to 50\% is about 300 ns.  The second mode is designated the ``Time-over-Threshold'' trigger (ToT) and at least 13 bins (i.e. $>$325 ns) in 120 FADC bins of a sliding window of 3$\mu$s are required to be above a threshold of 0.2 $I^{peak}_{VEM}$ in coincidence in 2 out of 3 PMTs\footnote{For detectors with only two (one) operating PMTs, the algorithm is applied to two (one) PMTs.}. This trigger is intended to select sequences of small signals spread in time.  
The ToT trigger is thus optimised for the detection of near-by, low energy showers, dominated by the electromagnetic component, or for high-energy showers where the core is distant.  The time spread arises from a combination of scattering (electromagnetic component) and geometrical effects (muons) as discussed in \cite{Linsley,Watson} where details are given of how the time spread depends on distance and zenith angle.  Since the average signal duration of a single muon is only about 150 ns, the time spread of the ToT (325 ns) is very efficient at eliminating the random muon background. The ToT rate at each detector is $<$ 2Hz and is mainly due to the occurrence of two muons arriving within 3$\mu$s, the duration of the sliding window.

The {\bf T2} is applied in the station controller to reduce to about 
20\,Hz the rate of events per detector. This reduction is done to cope with the
bandwidth of the communication system between the detectors and the central campus.
The T2 triggers, namely their time stamp and the kind of T2, are sent to the CDAS for the formation of the trigger of the array. 
All ToT-T1 triggers are promoted to the T2 level, whereas TH-T1 triggers are
requested to pass a further higher threshold of 3.2 $I^{peak}_{VEM}$ in coincidence
among the three PMTs\footnote{For detectors with only two (one) operating PMTs the threshold is set to 3.8 (4.5) $I^{peak}_{VEM}$.}.
The rates of the TH-T2 triggers are rather uniform in the detectors
over the whole array within a few percent, while those due to the ToT-T2 are less
uniform. This is due to the fact that the ToT is
very sensitive to the shape of the signal, this in turn depending
on the characteristics of the water, the reflective liner in the detector and the electronic pulse shaper. However, the lack of
uniformity of the trigger response over the array does not affect the event
selection or reconstruction above the energy corresponding to saturated acceptance.

\subsection{Trigger of the surface array }\label{t3}
The third level trigger, {\bf T3}, initiates the central data
acquisition from the array. It is formed at the CDAS, and it is based 
on the spatial and temporal combination of T2. Once a T3 is formed, all FADC signals from detectors passing the T2 are sent to the CDAS, as well as those from detectors passing the T1 but not the T2, provided that they are within 30 $\mu$s of the T3.

\begin{figure}[H]
\centering
\includegraphics[width=5cm]{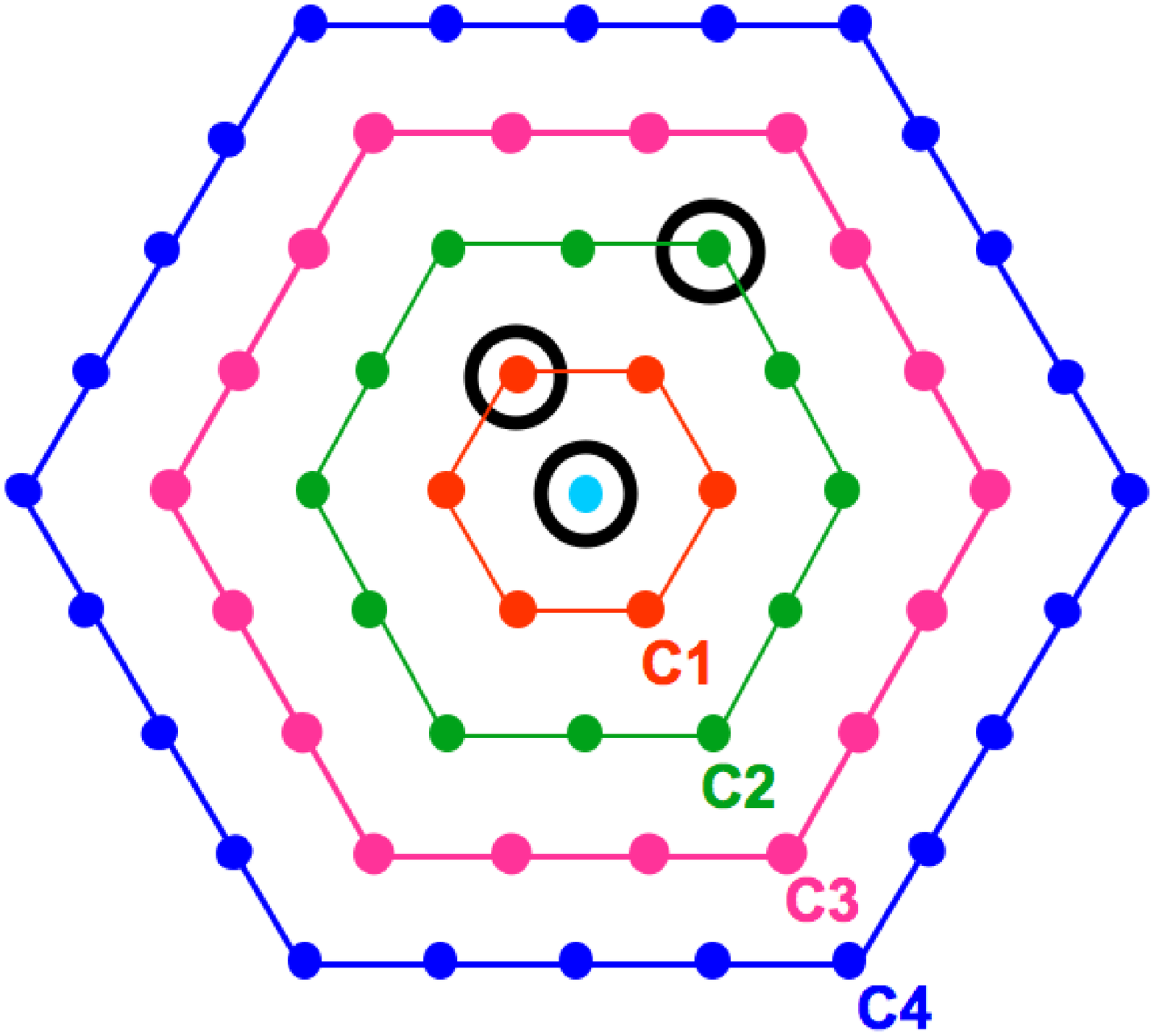}\hspace{1cm}\includegraphics[width=5cm]{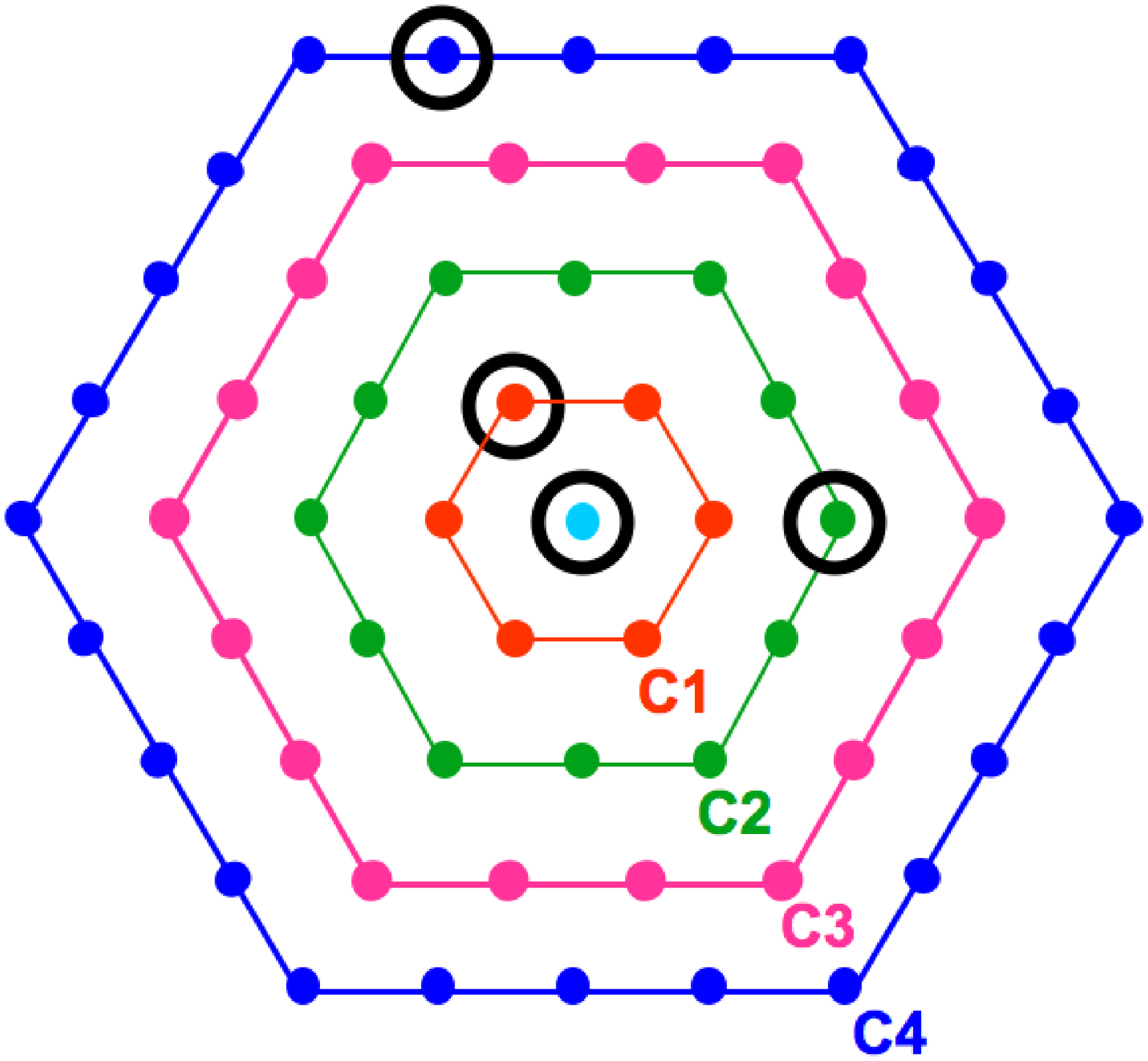}
\caption{Example of T3 configurations: the 3-fold T3 mode $ToT2C_1\&3C_2$ is shown on the left and the 4-fold mode $2C_1\&3C_2\&4C_4$ on the right (see text for the definitions). C1, C2, C3, C4 indicate the first, second, third and fourth sets of neighbours, respectively at 1.5, 3, 4.5 and 6 km from a given detector.}
\label{fig:t3}
\end{figure}

The trigger of the array is realised in two modes. The first T3 mode requires the coincidence of at least three detectors that have passed the
ToT condition and that meet the requirement of a minimum of compactness, namely, one of
the detectors must have one of its closest neighbours and one of its second closest
neighbours triggered. It is called "$ToT2C_1\&3C_2$", where $C_n$ indicates the $n^{th}$ set of neighbours (see figure \ref{fig:t3}). Once the spatial coincidence is verified, timing criteria are imposed: each T2 must be within $(6+5 C_{n}) \mu$s  of the first one.  An example of such T3 configuration is shown in figure \ref{fig:t3}, left. Since the ToT as a local trigger has very low background, this trigger selects predominantly
physics events. The rate of this T3 with the full array in operation
is around 1600 events per day, meaning
that each detector participates in an event about 3 times per day. This trigger
is extremely pure since 90\% of the selected events are real showers and it is
mostly efficient for showers below 60$^{\circ}$. The 10\% remaining are caused by chance coincidences due to the permissive timing criteria.
The second T3 mode is more permissive. It requires a four-fold coincidence
of any T2 with a moderate compactness. Namely, among the four fired detectors, within appropriate time windows,  at least one must be in the first set of neighbours from a selected station ($C_1$), another one must be in the second set ($C_2$) and the last one
can be as far as in the fourth set ($C_4$).  This trigger is called "$2C_1\&3C_2\&4C_4$". Concerning timing criteria,  we apply the same logic as for the "$ToT2C_1\&3C_2$". An example of such T3 configuration, is shown in figure \ref{fig:t3}, right. Such a
trigger is efficient for the detection of horizontal showers that, being rich in muons, generate in the detectors signals that have a narrow time spread, with triggered detectors having wide-spread patterns on the ground. With the full array configuration, this trigger selects about 1200 events per day, out of which about 10\% are real showers.

\subsection{Efficiency of the single detector trigger}

The single detector trigger probability as a function of the signal, $\mathcal{P}(S)$,
besides being important for the determination of the efficiency of the trigger of the array, is also of use in the event reconstruction where non-triggered detectors are included up to 10~km from a triggered one \cite{S1000rec}. 

\begin{figure}[H]
\centering
\includegraphics[width=11cm]{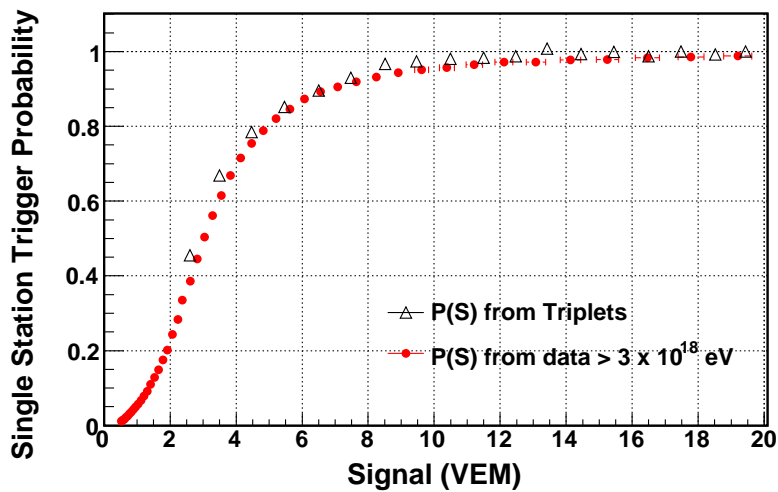}
\caption{Single detector trigger probability as a function of the signal in the detector, $\mathcal{P}(S)$, obtained from triplets data (triangles) and from showers data with $E > 3\times 10^{18}$ eV (circles).}\label{fig:pds}
\end{figure}

The T1 efficiency versus signal in the detector, $\mathcal{P}(S)$, is determined by using the very large statistics of EAS ($\approx 10^6$) recorded by the surface detector array. For each detected EAS, and each participating detector, we measure the trigger probability $\mathcal P(S)$  as the ratio $\frac{N_{T}(S)}{N_{ON}(S)}$, in different bins of $\uptheta$ and $S$(1000), of the number of triggered stations, $N_T$, to the total number of active stations, $N_{ON}$. $S$ is the $\emph{expected}$ signal at a detector, based upon the LDF fitted from the $\emph{measured}$ values from each detector, and $S$(1000) is the signal strength at 1 km, as derived from this fit. Since $\mathcal{P}(S)$ is obtained from events that actually produced a T3, the method is biased by events with a positive fluctuation in the signal. This bias can be corrected by Monte Carlo simulations and is found to be negligible at energies above around $3\times10^{18}$ eV. Limiting the analysis to showers with $S_{38}$ > 16 VEM (corresponding to about $3\times10^{18}$eV), the trigger probability versus signal is derived averaging over all the bins in $\uptheta$ and S(1000). This  is shown in figure~\ref{fig:pds} (circles):  the probability becomes $>0.95\%$ for $S\approx$ 10 VEM. This result is confirmed by an independent analysis that makes use of showers triggering certain detectors that have been specially located very close to one another. The surface array has seven positions in which three detectors (so called triplets) have been deployed at 11 m from each other. In each triplet, only one detector (master) sends T2 to CDAS, while the other two (slaves) are independently read out each time a T3 is generated and if they pass the T1.  For each slave, the trigger probability versus recorded signal $S$ is derived from the ratio between the number of events where both slaves have triggered and the number of events where only the other one has triggered. Depending if one or two slaves have triggered, $S$ is either the signal of the only triggered detector or the average of the two. 

From the analysis of about 10000 events, and combining the probabilities for the two slaves, $\mathcal{P}(S)$ is obtained and it is shown in figure~\ref{fig:pds} (triangles), in good agreement with the one obtained by showers data.

\section{Event selection of the surface detector array for showers with zenith angle below 60$^\circ$}
\label{sec:eventSelection}

A selection of physics events and of detectors belonging to
each event is made after data acquisition. Indeed, a large number of chance coincidence
events is expected due to the large number of possible combinations among the
single detectors. 
We focus here on the selection of events between 0$^\circ$ and 60$^\circ$ since more inclined showers have different properties and require specific selection criteria described elsewhere \cite{has}.

\begin{figure}[H]
\centering\includegraphics[width=14cm]{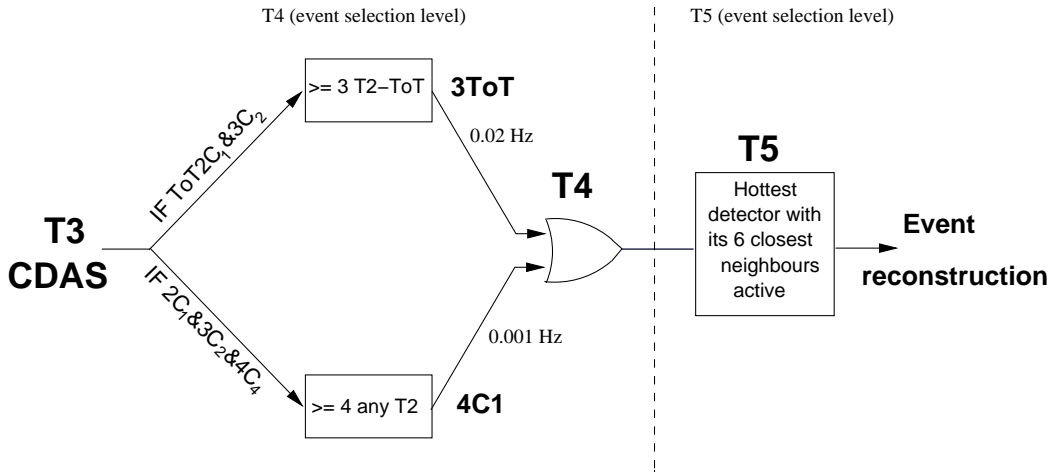}
\caption{Schematics of the hierarchy of the event selection of the Auger surface detector.}\label{fig:schema2}
\end{figure}

Two successive levels of selection are implemented. The first one
(physics trigger) is based on space and time configurations of the
detector, besides taking into account the kind of trigger in each of them. The
second one (fiducial trigger) requires that the shower selected by the physics trigger
is contained within the array boundaries, to guarantee the accuracy of the event reconstruction 
both in terms of arrival direction and energy determination. 
The logic of this off-line trigger system and its connection to the DAQ triggers  is summarised in figure~\ref{fig:schema2}. 

\subsection{Physics trigger}\label{t4}
The physics trigger, {\bf T4}, is needed to select real showers from the set of
stored T3 data. Two criteria are
defined, with different aims. The first T4 criterion, so-called 3ToT, requires 3 nearby stations, passing the T2-ToT, in a triangular pattern. It requires additionally that the times of the signals in the 3 stations fit to a plane shower front moving at the speed of the light. The number of chance coincidence passing the  3\,ToT condition over the full array is less than one per day, thanks to the very low rate of the T2-ToT.
Due to their compactness, events with zenith angles below 60$^\circ$are selected with high efficiency, i. e. more than $98\%$. 

The second T4 criterion, so called 4C1, requires 4 nearby stations, with no condition on the kind of T2. In this case also, it is required that the times of the signals in the 4 stations fit to a plane shower front moving at the speed of the light.
This 4C1 trigger brings to $\approx$100\% the efficiency for showers below 60$^\circ$.

\begin{figure}[H]
\centering\includegraphics[width=15cm]{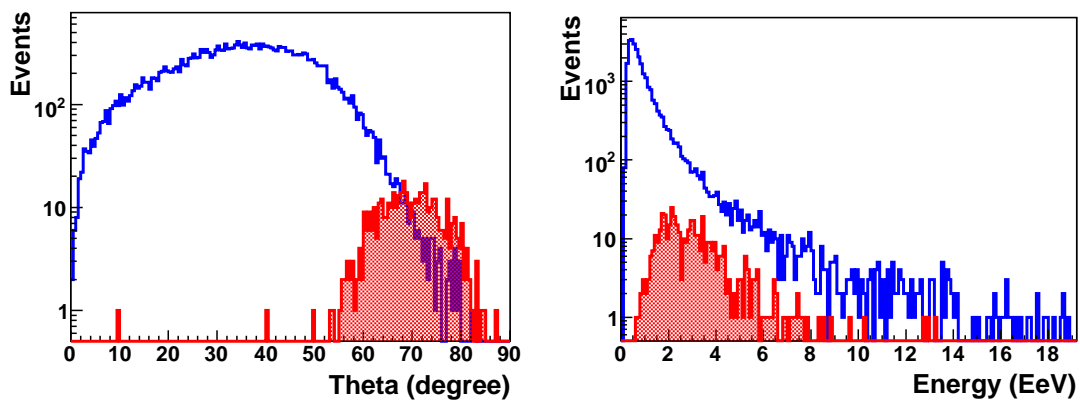}
\caption{Left: Angular (left) and energy (right) distribution of events selected by the T4 triggers: 3ToT (unfilled histogram), and 4C1, not ToT (filled histogram).}\label{fig:t4}
\end{figure}

The zenith angle distribution of events selected by the
T4 criteria is shown in figure \ref{fig:t4}, left, in the unfilled histogram for 3ToT, and in the filled one for the 4C1 that are not 3ToT: the two criteria are clearly complementary, the latter favouring the selection of events with larger zenith angles. In figure \ref{fig:t4}, right, the energy distributions of events selected by the two different criteria are shown: those selected by 3ToT have a median energy around $6\times 10^{17}$ eV, while for those selected by 4C1 it is around $3\times 10^{18}$ eV. 

Besides disentangling accidental events, there is also the need to identify, and reject, accidental detectors in real events, i.e. detectors whose signals are by chance in time with the others, but that in fact are not part of the event. To this aim, we define a "seed" made by 3 neighbouring detectors in a non-aligned configuration. If there is more than one triangle of stations, the seed with the highest total signal is chosen. If the T4 is a 3\,ToT, only ToT detectors can be considered to define the seed; if it is a 4C1, also TH detectors can be included.  Once the triangle has been determined, the arrival direction is estimated by fitting the arrival times of the signals to a plane shower front moving with the speed of light. Subsequently, all other detectors are
examined, and are defined as accidental if their time delay with respect to the front plane
is outside a time window of $[-2\,\mu\textrm{s}:+1\,\mu\textrm{s}]$. Detectors that have no triggered neighbours within 3\,km are always removed.

After the selection chain (both event selection and accidental detectors removal),
99.9\% of the selected events pass the full reconstruction procedure, that is arrival direction, core position and $S$(1000) are determined.

\subsection{Fiducial trigger}\label{t5}
\begin{figure}[H]
\centering
\includegraphics[width=8cm]{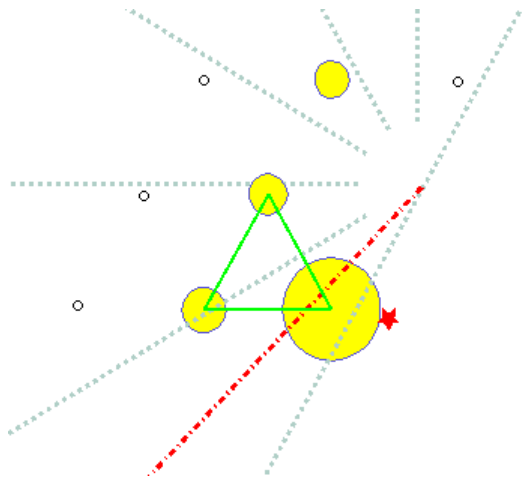}
\caption{Example of a hybrid, non-T5, event: the event falls on the border of the SD array, triggering only four detectors. Filled circles indicate the triggered ones, open circles the non-triggered active ones. The dimensions of the filled circles are proportional to the measured signal. The shower detector plane reconstructed by FD (dash-dotted line) indicates that the core is within the
triangle of detectors. The SD only reconstruction  places it outside the array (cross),
artificially increasing the event energy.}\label{fig:ste}
\end{figure}

The need for a {\it fiducial trigger}, {\bf{T5}}, mainly arises from events
falling close to the border of the array, where a part of the shower may be missing. In figure \ref{fig:ste} a hybrid event is shown, that triggered the SD and one of the FD telescopes, where a part of the SD information is missing due to its position on the border of the array.
Such events could have wrong core positions, and consequently, incorrect energies,
as in this example where the energy derived by SD is more than 4 times larger than the one estimated by FD ($1.4\times 10^{19}$ eV instead of $3\times 10^{18}$~eV). The main task of the fiducial trigger is thus to select only events well contained in the array, ensuring that the shower core is properly reconstructed. 

The fiducial trigger should be applied a priori on the events, to be independent of the reconstruction procedure. The T5 adopted requires that the detector with the highest signal has all its
6 closest neighbours working at the time of the event (i.e., it must be surrounded by
a working hexagon). 
This ensures adequate containment of the event inside the array. Even in
the case of a high energy event that falls inside, but close to the
border of the array, where part of the data may be missing, 
information from the seven detectors closest to the shower core ensures a proper reconstruction. Applying this condition, the maximum statistical uncertainty in the reconstructed S(1000) due to event sampling by the array is $\approx 3\%$ \cite{S1000rec}.
It has to be noted that this criterion also discards events that, though contained,
fall close to a non-working detector: this is an important issue because, due to the large number of detectors distributed over 3000 km$^2$, about 1\% of the detectors are expected to be not functioning at any moment, even with constant detector maintenance. For the fully completed array, and taking this into account, the application of the T5 condition reduces the effective area by 10\% with respect to the nominal one.

Finally, the use of the fiducial trigger allows the effective area of the array to
saturate to the geometrical one above a certain primary energy. Indeed, with no conditions on event containment, the acceptance would increase with increasing energy, since showers falling outside the borders of the array might still trigger sufficient detectors to be recorded; the higher their energy, the farther the distance.

\section{Aperture and exposure of the surface detector array for showers with zenith angle below 60 degrees}
The aperture of the surface detector array is given by the effective area integrated over solid angle. When the trigger and event selection have full efficiency, i.e. when the acceptance does not depend on the nature of the primary particle, its energy or arrival direction, the effective area coincides with the geometrical one.  
In subsection 5.1, the energy above which the acceptance saturates is derived. In section 5.2, the calculation of the exposure above this energy is detailed. 

\subsection{Determination of the acceptance saturation energy }\label{tiggereff}

\begin{figure}[H]
\begin{center}
\includegraphics[width=0.8\textwidth]{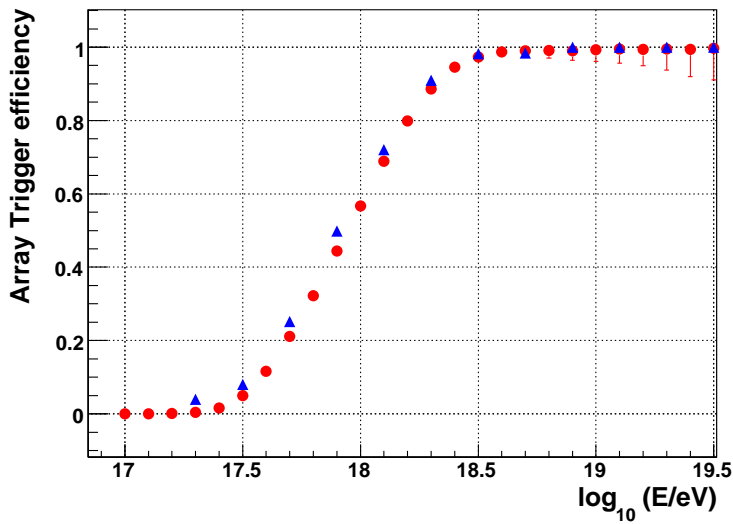}
\caption{Trigger efficiency as a function of energy, derived from SD data (triangles) and hybrid data (circles). } 
\label{fig:efftrigger}
\end{center}
\end{figure}

{\bf{I. From SD data}}. The acceptance saturation energy, $E_{SAT}$, is determined using two different methods that use events recorded by the surface detector array. In the first one, starting from detected showers, mock events are generated by fluctuating the amplitude of the signals recorded in each detector and their arrival time. Such fluctuations are measured \cite{signal,time} by using twin detectors located at 11~m from each other. To each simulated event, the full trigger and event selection chain are applied. From the ratio of the number of triggered events to the simulated, the trigger efficiency is obtained as a function of energy, as shown in figure \ref{fig:efftrigger} (triangles).  As can be seen, the trigger probability becomes almost unity ($>97\%$) at energy $E \sim 3 \times 10^{18}$ eV for all angles between 0$^\circ$ and 60$^{\circ}$. The fact that the method is based on the use of showers that actually triggered the array may bias the estimation of the trigger probability at low energy. However, it  does not bias the result on the trigger probability close to full efficiency, and hence on $E_{SAT}$. 

{\bf{II. From hybrid data}}. The hybrid data sample is composed of events observed by the FD and that triggered at least one SD detector: consequently, it has an intrinsically lower energy threshold than the SD. For each bin in energy (of width 0.2 in $\log_{10}$(E)), the number of events that pass the SD trigger out of the total number of events are counted. To avoid biases from primary composition, the same data selection criteria as in \cite{ElongationRate} are used. Additionally, in analogy with the T5,  to avoid the effects of the borders of the array, it is required that the detector used in the hybrid geometry reconstruction is surrounded by 6 active detectors. The trigger efficiency of the surface detector array is found to be saturated ($>$ 97\% ) for energies above $3\times10^{18}$ eV, as shown in figure \ref{fig:efftrigger} (circles), in agreement with what is obtained by the analysis of SD data alone.

\begin{figure}[H]  
\begin{center}
\includegraphics[width=0.8\textwidth]{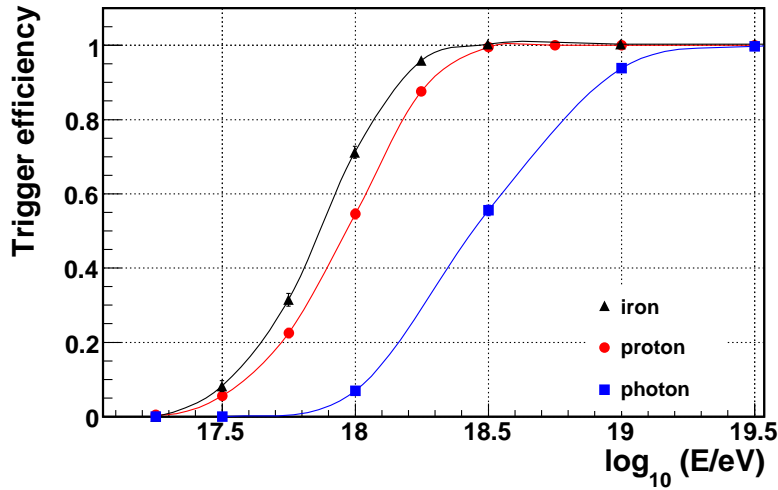}
\caption{SD trigger efficiency as a function of Monte Carlo energy $E$ for proton (circles), iron (triangles) and photon primaries (squares) and zenith angle integrated up to 60$^{\circ}$. Lines are drawn only to guide the eyes.} 
    \label{fig:3tot}
  \end{center}
\end{figure}

{\bf{III. Cross-check with simulations}}. $E_{SAT}$ is finally cross-checked using full shower and detector simulations. The simulation sample consists of about 5000 proton, 5000 photon and 3000 iron showers simulated using CORSIKA~\cite{Corsika} with zenith angle distributed as $\sin\uptheta\cos\uptheta$ ($\uptheta < $60$^{\circ}$) and energies ranging between 10$^{17}$~eV and 10$^{19.5}$~eV in steps of 0.25 (0.5 for photons) in $\log_{10}$(E).
The showers are generated using QGSJET-II~\cite{qgsjetII} and FLUKA \cite{fluka} for high and low energy hadronic interactions, respectively. Core positions are uniformly distributed at ground and each shower is used five times, each time with a different core position, to increase the statistics with a negligible degree of correlation. The surface detector array response is simulated using Geant4~\cite{geant4} within the framework provided by the \Offline software~\cite{offline}. The resulting trigger probability as a function of the Monte Carlo energy for proton, iron and photon primaries is shown in Figure~\ref{fig:3tot} for $0^\circ<\uptheta<60^\circ$. Due to their larger muon content, at low energies iron primaries are slightly more efficient at triggering the array than protons. However, the trigger becomes fully efficient at $3\times10^{18}$ eV, both for proton and iron primaries, in different intervals of zenith angles. It is important to notice that the trigger efficiency for photons is much lower. This is because photons tend to produce deeper showers that are poor in muons.

\subsection{Calculation of the integrated exposure} \label{sec:integratedExposure}

The studies described above have shown that the full efficiency of the SD trigger and event selection is reached at $3 \times10^{18}$~eV. Above this energy, the calculation of the exposure is based solely on the determination of the geometrical aperture and of the observation time.

\begin{figure}[H]  
\begin{center}
\includegraphics[width=0.3\textwidth]{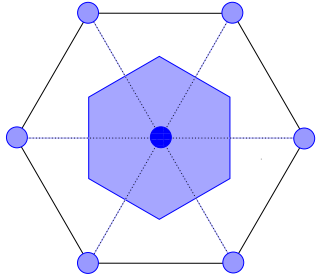}
\caption{Scheme of an hexagon of detectors: the elemental hexagon cell, $a_{\mathrm{cell}}$, is the shaded area around the central detector. } 
 \label{fig:hexa}
  \end{center}
\end{figure}

With respect to the aperture, the choice of a fiducial trigger based on hexagons, as explained in section~\ref{t5}, allows us to exploit the regularity of the array very simply. The aperture of the array is obtained as a multiple of the aperture of an elemental hexagon cell, $a_{\mathrm{cell}}$, defined as any active detector with six active neighbours, as shown in figure~\ref{fig:hexa}. 

\begin{figure}[H]  
\begin{center}
\includegraphics[width=0.8\textwidth]{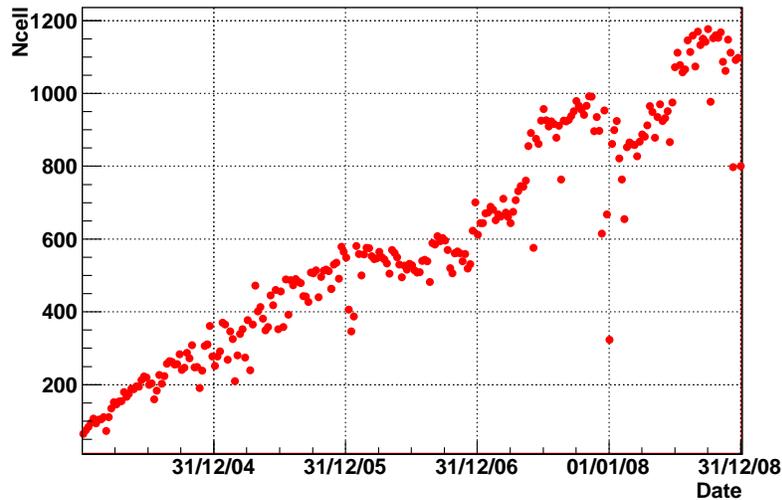}
\caption{Evolution of the number of hexagonal cells (see text) between January 1$^{st}$, 2004 and December 31$^{st}$, 2008} 
    \label{fig:cells}
  \end{center}
\end{figure}

At full efficiency, the detection area per cell is $1.95\,\mathrm{km}^{2}$. The corresponding aperture for showers with $\uptheta < 60^{\circ}$ is then $a_{\mathrm{cell}} \simeq 4.59\,\mathrm{km}^{2}\,\mathrm{sr}$.  The number of cells, $N_{\mathrm{cell}}(t)$, is not constant over time due to temporary problems at the detectors (e. g. failures of electronics, power supply, communication system, etc...).  $N_{\mathrm{cell}}(t)$ is monitored second by second: we show in figure  \ref{fig:cells}  the evolution of $N_{\mathrm{cell}}(t)$ between the start of the data taking, January 2004, and December 2008. Such precise monitoring of the array configurations allows us to exploit data during all deployment phases, clearly visible in the figure, as well as during unstable periods as during, for example, January 2008 when huge storms affected the communication system.
The second-by-second monitoring provides at the same time the aperture of the array per second,  $a_{\mathrm{cell}} \times N_{\mathrm{cell}}(t)$, as well as the observation time with high precision. To calculate the integrated exposure over a given period of time, the aperture of the array, $N_{cell}(t) \times a_{cell}$, is integrated over the number of live seconds. This calculation is expected to be very precise, since it is based on a purely geometrical aperture and a very good time precision. However both the determination of $N_{\mathrm{cell}}(t)$ and of the observation time are affected by uncertainties.

Concerning the determination of $N_{\mathrm{cell}}(t)$, to evaluate the uncertainty in the number of active detectors, a check of the consistency of the event rate of each detector with its running time, determined from the monitoring system, is performed. The uncertainty derived from this study is added to that due to errors of communication between the station and the DAQ, which are also monitored. Overall, the uncertainty on the determination of $N_{\mathrm{cell}}(t)$ amounts to about 1.5\%.

\begin{figure}[H]  
\begin{center}
\includegraphics[width=0.8\textwidth]{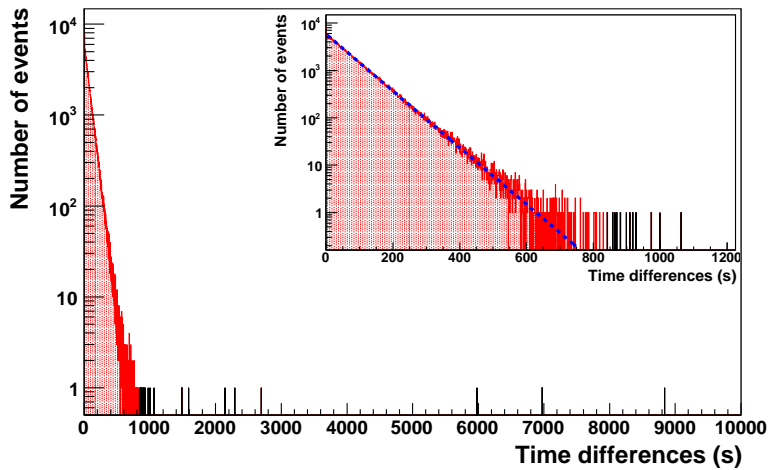}
\caption{Distribution of time differences between events in  2008. The points outside the filled area show the dead times (see text). The exponential fit is shown as a dashed line in the inset where the histograms are zoomed. }
  \label{fig:deltat}
  \end{center}
\end{figure}

For the determination of the observation time, and related uncertainty, the dead time that is unaccounted for in the second by second monitoring of the array, is taken into account\footnote{This dead time can be due either to problems in the communication between the stations and the CDAS or to problems of data storage in the stations  }. To determine these, an empirical technique is exploited, based on the study of the distribution of the arrival times of events, under the reasonable hypothesis that they follow a Poisson distribution. Given the constant rate $\lambda$ for the T5 event rate per hexagon, $\lambda \approx 1.4 \times10^{-5}$ event per second per hexagon, the probability $P$ that the time interval T
between two consecutive T5 events be larger than T is given by:  $P (T ) = e^{-\lambda T}$. We define intervals as dead time if the Poisson probability of their occurrance is less than $10^{-5}$.  As an example, we show in figure \ref{fig:deltat} the distribution of time differences for events acquired in 2008. The distribution is exponential with a time constant of 72.4 seconds, as expected for the above value of $\lambda$ and the observed average number of live hexagons during that year.  In the figure, the points outside the filled area show those time intervals that have occurred with a Poisson probability less than $10^{-5}$.
The identified dead times generally correspond to periods of software modifications at the level either of the single detectors or of the CDAS. These were rather frequent during the deployment phase of the surface detector array, which lasted until June 2008. The uncertainty in the determination of the livetime is estimated to be around 1\%. Between January 2004 and December 2008, the livetime of the surface detector array data acquisition is 96\%. Hidden dead times reduce the effective livetime to 87\%, the reduction being mostly due to the two first years of operation. However, due to the growth of the surface detector array,  their impact on the total integrated exposure is a reduction of only 3\%.

\section{Conclusions}

The DAQ trigger of the surface detector array of the Pierre Auger Observatory is organised in a hierarchical way, starting at the level of the single detector (T1, T2) up to the data acquisition (T3). The selection of events below 60$^{\circ}$ takes place off-line, and it is also hierarchical (T4, T5). The whole chain, from the single detector trigger, up to event selection, is able to reduce the counting rate of the single detector from about 3 kHz, due mainly to single, uncorrelated, cosmic muons, down to about $3\times 10^{-5}$ Hz. This final rate is due to extensive air showers, more than 99\% of which pass the reconstruction chain.

In spite of the large number of detectors and the possible number of chance events due to combinatorial coincidences among the detectors, the high-purity Time Over Threshold trigger enables the main trigger of the array to be kept at the level of a 3-fold coincidence, thus extending the range of physics that can be studied. Such a trigger, together with the event selection strategy, allows the acquisition and reconstruction of about one cosmic ray shower per minute, with median energy around $6\times~10^{17}$~eV. Moreover, it makes the surface detector array fully efficient for showers due to primary cosmic rays above $3\times10^{18}$~eV, independent of their mass and arrival directions. The trigger provides at the same time a larger overlapping energy region with the FD, which is naturally efficient at lower energies, allowing the measurement of the cosmic ray spectrum down to $10^{18}$ eV \cite{Fabian}.

\begin{figure}[H]  
\begin{center}
\includegraphics[width=0.8\textwidth]{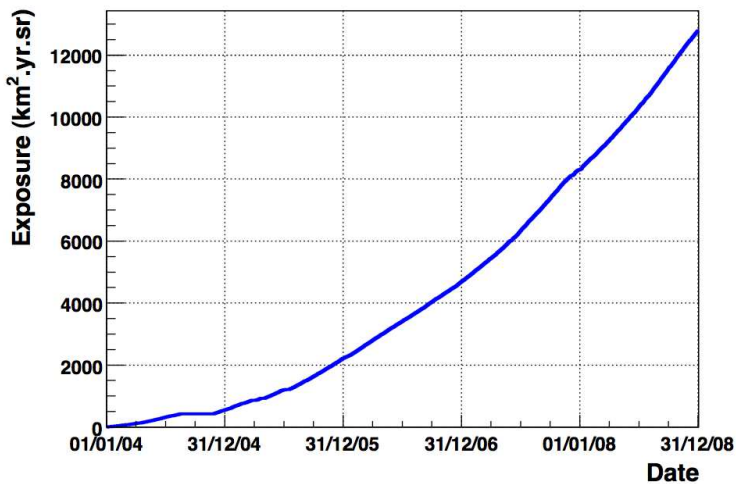}
\caption{Evolution of the integrated exposure between January 1$^{st}$, 2004 and December 31$^{st}$, 2008} 
    \label{fig:expo}
  \end{center}
\end{figure}

Above $3 \times 10^{18}$ eV, the calculation of the exposure is purely geometrical, being the integration of the geometrical aperture over the observation time. Both of them are known with high precision, so that the overall uncertainty on the integrated exposure is less than 3\%. The integrated SD exposure as a function of time is shown in figure \ref{fig:expo}, from January 2004 to December 2008: at the end of the period it amounts to $12790\pm380$~km$^{2}$\,sr\,yr. Even though the SD was under continuous deployment until June 2008, the effective livetime of the surface detector array averaged over all the five years is high, being 87\%. The effective livetime of the SD is 96\% for 2008 alone: with this livetime and the full surface detector array deployed, the exposure is expected to increase by about 500 km$^2$\,sr\,yr per month.

\section{Acknowledgements}

The successful installation and commissioning of the Pierre Auger Observatory
would not have been possible without the strong commitment and effort
from the technical and administrative staff in Malarg\"ue.

We are very grateful to the following agencies and organizations for financial support: 
Comisi\'on Nacional de Energ\'ia At\'omica, 
Fundaci\'on Antorchas,
Gobierno De La Provincia de Mendoza, 
Municipalidad de Malarg\"ue,
NDM Holdings and Valle Las Le\~nas, in gratitude for their continuing
cooperation over land access, Argentina; 
the Australian Research Council;
Conselho Nacional de Desenvolvimento Cient\'ifico e Tecnol\'ogico (CNPq),
Financiadora de Estudos e Projetos (FINEP),
Funda\c{c}\~ao de Amparo \`a Pesquisa do Estado de Rio de Janeiro (FAPERJ),
Funda\c{c}\~ao de Amparo \`a Pesquisa do Estado de S\~ao Paulo (FAPESP),
Minist\'erio de Ci\^{e}ncia e Tecnologia (MCT), Brazil;
AVCR AV0Z10100502 and AV0Z10100522,
GAAV KJB300100801 and KJB100100904,
MSMT-CR LA08016, LC527, 1M06002, and MSM00216\-20859, Czech Republic;
Centre de Calcul IN2P3/CNRS, 
Centre National de la Recherche Scientifique (CNRS),
Conseil R\'egional Ile-de-France,
D\'epartement  Physique Nucl\'eaire et Corpusculaire (PNC-IN2P3/CNRS),
D\'epartement Sciences de l'Univers (SDU-INSU/CNRS), France;
Bundesministerium f\"ur Bildung und Forschung (BMBF),
Deutsche Forschungsgemeinschaft (DFG),
Finanzministerium Baden-W\"urttemberg,
Helmholtz-Gemein\-schaft Deutscher Forschungszentren (HGF),
Ministerium f\"ur Wissenschaft und Forschung, Nordrhein-Westfalen,
Ministerium f\"ur Wissenschaft, For\-schung und Kunst, Baden-W\"urttemberg, Germany; 
Istituto Nazionale di Fisica Nucleare (INFN),
Ministero dell'Istruzione, dell'Universit\`a e della Ri\-cerca (MIUR), Italy;
Consejo Nacional de Ciencia y Tecnolog\'ia (CONACYT), Mexico;
Ministerie van Onderwijs, Cultuur en Wetenschap,
Nederlandse Organisatie voor Wetenschappelijk Onderzoek (NWO),
Stichting voor Fundamenteel Onderzoek der Materie (FOM), Netherlands;
Ministry of Science and Higher Education,
Grant Nos. 1 P03 D 014 30, N202 090 31/0623, and PAP/218/2006, Poland;
Funda\c{c}\~ao para a Ci\^{e}ncia e a Tecnologia, Portugal;
Ministry for Higher Education, Science, and Technology,
Slovenian Research Agency, Slovenia;
Comunidad de Madrid, 
Consejer\'ia de Educaci\'on de la Comunidad de Castilla La Mancha, 
FEDER funds, 
Ministerio de Ciencia e Innovaci\'on,
Xunta de Galicia, Spain;
Science and Technology Facilities Council, United Kingdom;
Department of Energy, Contract No. DE-AC02-07CH11359,
National Science Foundation, Grant No. 0450696,
The Grainger Foundation USA; 
ALFA-EC / HELEN,
European Union 6th Framework Program,
Grant No. MEIF-CT-2005-025057, 
European Union 7th Framework Program, Grant No. PIEF-GA-2008-220240,
and UNESCO.

\end{document}